\documentclass[aps,pre,twocolumn,showpacs]{revtex4}
\usepackage{graphicx}
\begin{document}
\title{Centrality Scaling of the $p_T$ Distribution of Pions}
\author{Rudolph C. Hwa$^1$ and C.\ B.\ Yang$^{1,2}$}
\affiliation{$^1$Institute of Theoretical Science and Department of Physics\\
University of Oregon, Eugene, OR 97403-5203, USA\\
$^2$Institute of Particle Physics, Hua-Zhong Normal University, Wuhan
430079, China}

\begin{abstract}
From the preliminary data of PHENIX on the centrality dependence of
the $\pi^0$ spectrum in $p_T$ at midrapidity in heavy-ion collisions,
we show that a scaling behavior exists that is independent of the
centrality. It is then shown that $\left<p_T\right>$ degrades with
increasing $N_{\rm part}$ exponentially with a decay constant that
can be quantified. A scaling distribution in terms of an intuitive
scaling variable is derived that is analogous to the KNO scaling. No
theoretical models are used in any part of this phenomenological
analysis.

\end{abstract}
\maketitle

In a recent paper \cite{hy} we reported on the finding of a scaling
property of the $p_T$ distribution of pions produced in heavy-ion collisions
that is independent of the collision energy. Here we present an
extension of that scaling property to include centrality variations
and show that a KNO-type \cite{kno} scaling behavior exists over the
entire range of $p_T$ measured. The investigation is primarily a
phenomenological analysis with no assumptions about the hard and soft
collisions, nor about the parton energy losses.

Recently, a scaling behavior of the transverse-mass spectrum has been
reported in \cite{jsb}. That work was motivated by color glass condensate and the
saturation of the gluon density in nuclear collisions. Our investigation has no theoretical
motivation other than the search for the simplest form that can
represent the data. The dynamical origin of the $p_T$ distribution is
complicated. At low $p_T$ the statistical model seems to work well,
as does the hydrodynamical description up to $p_T=$3 GeV/c \cite{ph}.
At high $p_T$ hard parton scattering will create jets, which can lose
energy  due to multiple scatterings of partons in the dense medium
\cite{xnw}. A universal description of the hadron distribution over
all $p_T$ is nonexistent, if not meaningless from the point of view
of the sectarian nature of the dynamical theories that claim
validities in different domains. However, if a universal scaling
behavior can be found phenomenologically, it can serve as a common
goal for different dynamical approaches to aim at.
\begin{figure}[tbph]
\includegraphics[width=0.5\textwidth]{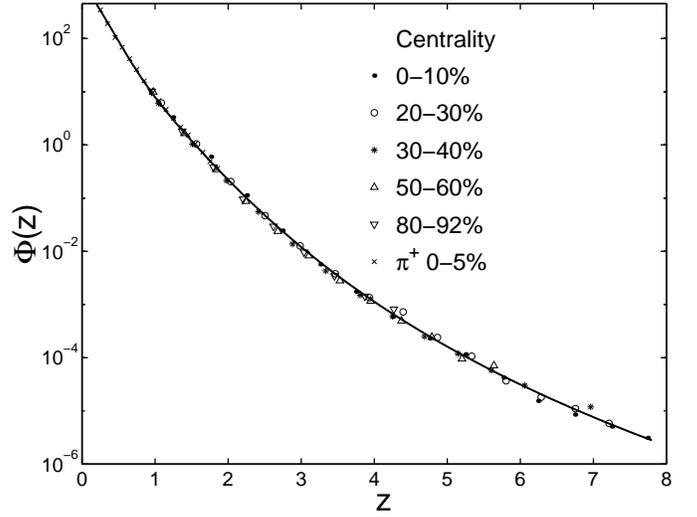}
\caption{Scaling distribution $\Phi(z)$ showing the coalescence of
5 centrality bins of the preliminary data from PHENIX on $\pi^0$
production in Au+Au collisions at $\sqrt s=200$ GeV \cite{dde}.
The points labeled by $\times$ are for $\pi^+$.
The solid line is a fit parametrized by Eq.\ (\ref{5}).}
\end{figure}

From the preliminary PHENIX data of $\pi^0$ produced in Au+Au
collisions at the relativistic heavy-ion collider (RHIC) \cite{dde},
we have the pion distribution,
$(2{\pi}p_T)^{-1}dN_{\pi}/dp_T$, at midrapidity for $\sqrt s = 200$
GeV and for a wide range of centrality that has 9 bins from 0-10\% to
80-92\%. To unify the 9 distributions, it is necessary to define a
scaling variable $z$. First, we use the number of participants,
$N_{\rm part}$, to quantify centrality; those numbers for different
bins are taken from \cite{kn}, which agree well with those given by
PHENIX \cite{ad}.  Next, we define, for fixed $\sqrt s$ (at 200 GeV),
\begin{equation}
z=p_T / K(N)    ,     \label{1}
\end{equation}
where $K$ depends on $N_{\rm part}$, for which we use the abbreviated notation
$N=N_{\rm part}$ hereafter. For every centrality bin we vary $K$
by plotting the data of $(2{\pi}p_T)^{-1}dN_{\pi}/dp_T$ in terms of
$z$ and adjusting the normalization so that all data points lie on a
universal curve. That is, we define
\begin{equation}
\Phi(z)=A(N)\,K^2(N)\,{1\over 2{\pi}p_T}{dN_\pi\over dp_T},
\label{2}
\end{equation}
and find $A(N)$ and $K(N)$ such that $\Phi(z)$ has no
explicit dependence on $N$. That turns out to be possible, as
evidenced by Fig.\ 1. For clarity we show only 5 bins of centrality
in that figure. It is a remarkable  property of the centrality
dependence of the pion spectra that such a universal scaling
distribution exists.
\begin{figure}[tbph]
\includegraphics[width=0.45\textwidth]{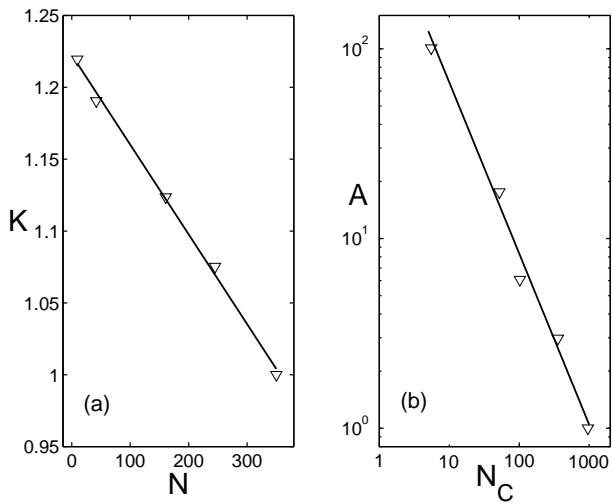}
\caption{(a) Scale factor $K(N)$ in units of GeV/c. Solid line is
a fit by Eq.\ (\ref{3}).
(b) Power-law behavior of the normalization factor $A(N_c)$.
Solid line is a fit by Eq.\ (\ref{4}).}
\end{figure}

The values of $K(N)$ used to obtain the scaling behavior are shown
in Fig.\ 2(a) in units of GeV/c. The dependence of
$K(N)$ on $N$ can be well fitted by
\begin{equation}
K(N)=1.226 - 6.36{\times}10^{-4} N,     \label{3}
\end{equation}
such that  $K(N_{\rm max})=1$ at $N=N_{\rm max}=350$. The effects of
the degradation of parton momenta are hidden in this formula. Any
change of the overall scale of $K(N)$ is trivial and does not affect
the scaling behavior that we have found. Although the normalization factor
$A(N)$ does not have a simple dependence on $N$, it turns out to
depend simply on the number of binary collisions $N_c$. The values of
$A(N_c)$ needed to achieve the scaling $\Phi(z)$ are shown in Fig.\ 2(b) in a log-log plot.
They can be fitted by
\begin{equation}
A(N_c) = 530 N_c^{-0.9}.   \label{4}
\end{equation}
From the tables listed in Refs.\ \cite{kn,ad}, $N_c$ and $N$ can be related by
 $N_c=0.44 N^{1.33}$. Note that the normalization of $\Phi(z)$ is set by the
most central bin by choosing $A(N)=1$ at $N=N_{\rm max}$. If $A(N_c)$
were to behave as $N_c^{-1}$, it would suggest that the average
multiplicity of pions at midrapidity is proportional to $N_c$, which
is a variable that measures the number of hard collisions. Thus the
factor $N_c^{-0.9}$ in Eq.(\ref{4}) is an indication that the
centrality dependence of the midrapidity multiplicity scales as
$N_c^{0.9}$ from the $pp$ collisions, revealing the effect of
suppression of $p_T$ in the nuclear medium.

To fit the scaling curve the $\pi^0$ data are insufficient to give us
guidance in the small $z$ region, since they do not extend below $p_T$=1
GeV/c. For $0<p_T<1$ GeV/c, we use the $\pi^+$ data of PHENIX for 0-5\%
centrality \cite{ch} shown in Fig.\ 1. The combined $\pi^0$ and $\pi^+$
data can be well fitted by
\begin{equation}
\Phi(z)=1200\,(z^2 + 2)^{-4.8} (1+25\,e^{-4.5\,z}),     \label{5}
\end{equation}
which is shown by the solid line in Fig.\ 1. We can check its
normalization by evaluating the integral
\begin{equation}
I=\int_0^{10} dz\,z\,\Phi(z)=46.2={A(N)\over 2\pi} {dN_{\pi^0}\over
d\eta}.      \label{6}
\end{equation}
For $N=200$, say, this gives $dN_{\pi^0}/d\eta=149$, which compares
satisfactorily to $dN_{\rm ch}/d\eta/(0.5N)=3.2$ at the same $N$
\cite{ad}. Since the $\pi^{\pm}$ data do not extend into the $p_T>2$
GeV/c region, we do not consider them for centrality analysis here.

The exponential term in Eq.\ (\ref{5}) is mainly to fit the low-$z$ data
that contain thermodynamical effects. At high $z$, $\Phi(z)$ behaves as
a power law that represents the effects of hard collisions and jet
quenching. For all $z$, $\Phi(z)$ is a succinct summary of all dynamical
effects for all centralities.

In terms of $\Phi(z)$ it is now possible to have an analytic
expression of the inclusive distribution of the pions in $p_T$ at
midrapidity. For convenience, we shall write it in terms of the
momentum fraction $x$
\begin{equation} x=p_T / K_0,    \label{7}
\end{equation} where $K_0$ is a fixed scale, beyond which no physics
of interest need be of concern here. We set $K_0=10$ GeV/c for now,
although increasing it later, if necessary, is a simple matter. In
view of Eq.\ (\ref{1}) we thus have
\begin{equation} z=x \Lambda(N),   \qquad\qquad
\Lambda(N)=K_0/K(N).     \label{8}
\end{equation}
Converting $(2{\pi}z)^{-1}dN_{\pi}/dz$ to the $x$
variable, we define the corresponding pion distribution to be
\begin{equation}
H(x,N)=A^{-1}(N)\,\Phi(x,N),         \label{9}
\end{equation}
where $A(N)=A(N_c(N))$. To see the evolution of the pion distribution with
increasing $N$, it is more enlightening to study the normalized
distribution, defined by
\begin{equation}
P(x,N)=H(x,N)\left/\int_0^1 dx\,x\,H(x,N)\right.,        \label{10}
\end{equation}
where the upper limit of integration is set to 1 on the assumption that the
contribution from
$p_T>K_0$ is  insignificant. Thus
$P(x,N)$ is the probability distribution of producing a $\pi^0$ at
$x$, for which the
differential phase space is $xdx$ due to the 2D nature of $\vec p_T$.
\begin{figure}[tbph]
\includegraphics[width=0.45\textwidth]{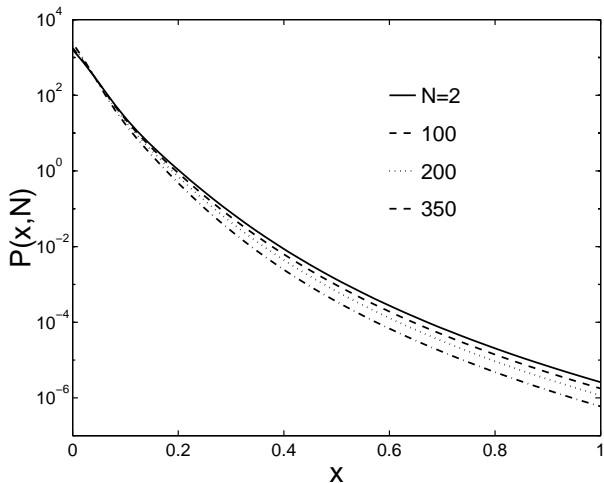}
\caption{Probability distribution $P(x,N)$ for 4 values of $N$.}
\end{figure}

In Fig.\ 3 we show $P(x,N)$ for 4 values of $N$. Note how $P(x,N)$
decreases at high
$x$ but increases at low $x$, when $N$ is increased. That is the
behavior we expect when high-$p_T$ partons are suppressed, giving
rise to low-$p_T$ partons. The crossover occurs at around $x=0.06$,
corresponding to $p_T=0.6$ GeV/c.

Such an evolution of the $x$-distribution is reminiscent of the
evolution of the parton distribution in ln$\,Q^2$ in perturbative
QCD. Although no precise relationship between the two has been
established, it is known that in the latter case the analytical
description is simpler in terms of the moments. Thus let us
define the moments
\begin{equation}
P_n(N)=\int_0^1 dx\,x^{n+1}\,P(x,N).     \label{11}
\end{equation}
From Eqs.\ (\ref{5}), (\ref{9}) and (\ref{10}) we can
calculate the $N$ dependencies of $P_n(N)$, which are shown in Fig.\
4 for $n=1$ to 5. Evidently, ln$\,P_n(N)$ can be well approximated by
linear dependence on $N$, i.e.,
\begin{equation}
\ln P_n(N)=a_n-b_n\,N.     \label{12}
\end{equation}
The slope parameters $b_n$ are shown in the inset of the same
figure. The dependence of $b_n$ on $n$ is also linear. Thus we may
rewrite Eq.\ (\ref{12}) as
\begin{equation}
{d\over dN} \ln P_n(N) = -\lambda\ n,  \qquad
\lambda=5.542\times 10^{-4}.
\label{13}
\end{equation}
This is a very economical way of describing  the
degradation property of the pion distribution in terms of one basic
parameter $\lambda$.

A physical interpretation can readily be given for $\lambda$ when we
consider $n=1$, for which $P_1(N)=\left< x\right>_N$, the average $x$ at $N$.
From Eq.\ (\ref{13}) we obtain
\begin{equation}
\left< x\right>_N=\left< x\right>_{N_0}\,\exp\,[-\lambda\,(N-N_0)],
\label{14}
\end{equation}
which exhibits explicitly the exponential decrease of
$\left<x\right>_N$ with increasing $N$, a behavior that solidifies
our physical notion of what the dense medium does to
$\left<p_T\right>$. For $N_0=2$ and $N=N_{\rm max}$, we get
\begin{equation}
\left<x\right>_{N_{\rm max}}\,/\,\left<x\right>_2 = 0.825,
\label{15}
\end{equation}
which gives a quantitative measure of the degree of degradation.
From Eq.\,(\ref{13}) it is easy also to show that
\begin{equation}
{d\over dN}\,{\left<x^n\right>_N\over\left<x\right>_N^n} = 0,
\label{16}
\end{equation}
where $\left<x^n\right>_N=P_n(N)$. Hence, the
normalized moments of $P(x,N)$ are invariant in $N$. That is a clue
to another invariant form of the distribution.
\begin{figure}[tbph]
\includegraphics[width=0.45\textwidth]{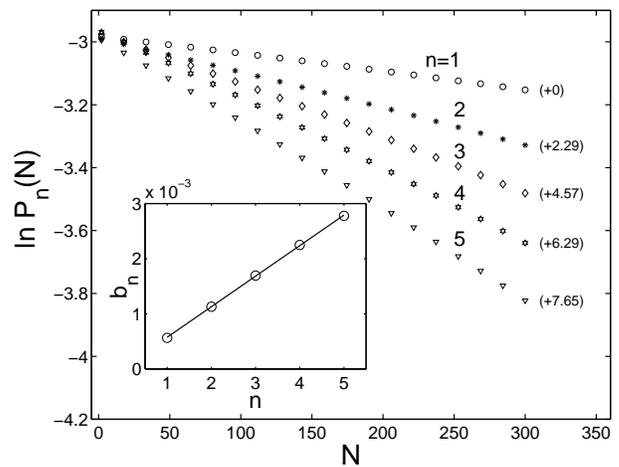}
\caption{$N$ dependence of the moments, $P_n(N)$, whose log values are raised
by the quantities in the parentheses.  The inset shows
the slopes $b_n$, the line being a linear fit.}
\end{figure}

Before we examine the implications of that clue, we note that the
properties of $P_n(N)$ displayed in Fig.\ 4 and described by Eq.\ (\ref{12}) cannot
be expected to be valid for arbitrarily large n, since the definition
of $P_n(N)$ in Eq.\ (\ref{11}) puts more weight on the high end of
$x$ when $n$ is large. Our cutoff at $x=1$, corresponding to
$p_T=K_0=10$ GeV/c, is based partly on the lack of data at higher
$p_T$ and partly on the recognition that the contribution from
$p_T>K_0$ is unimportant when $n$ is not too large. To test the
validity of our procedure, we have carried out the analysis for
$K_0=20$ GeV/c, using the same $\Phi(z)$, and found that Eq.\
(\ref{13}) remains to be an excellent approximation of the $n$
dependence  shown in Fig.\ 4, and that the value of $\lambda$ is
larger by just 2\%, which is less than the experimental errors. Thus
we claim that our analysis is stable under variations of $K_0$ so
long as we consider $K_0\ge10$ GeV/c and $n\le5$.

The invariance of the normalized moments in Eq.\ (\ref{16}) suggests
that we should consider yet another scaling variable
\begin{equation}
u=x\,/\left<x\right>_N=p_T\,/\left<p_T\right>_N         \label{17}
\end{equation}
for any fixed $N$. Let us now define
\begin{equation}
\Psi(u,N)=\left<x\right>_N^2\ P(x,N),     \label{18}
\end{equation}
whose moments are defined by
\begin{equation}
\Psi_n(N)=\int_0^{\left<x\right>_N^{-1}} du\ u^{n+1}\ \Psi(u,N).
\label{19}
\end{equation}
Transforming this integral to an integration over $x$, we find that
\begin{equation}
\Psi_n(N)=\left<x\right>_N^{-n}\,P_n(N).       \label{20}
\end{equation}
It then follows from Eq.\ (\ref{16}), that
\begin{equation}
d\,\Psi_n(N)\,/dN=0.      \label{21}
\end{equation}
Hence, $\Psi_n(N)$ is independent of $N$ and we have a
scaling function $\Psi_n$, which in turn implies that $\Psi(u)$ is
also independent of $N$. Indeed, from Eqs.\ (\ref{9}) and (\ref{10})
we see that (\ref{18}) can be reexpressed as
\begin{equation}
\Psi(u)=\Phi(z(u))\,\left/\int du\,u\,\Phi(z(u))\right.,
\label{22}
\end{equation}
where, by virtue of Eqs.\ (\ref{8}) and (\ref{17}),
\begin{equation}
z(u)=\left<x\right>_N\Lambda(N)\ u.     \label{23}
\end{equation}
Although $\left<x\right>_N\Lambda(N)$ may appear to
depend on $N$, it actually is a constant
\begin{equation}
\gamma=\left<x\right>_N\Lambda(N)=\left<z\right> ={\int
dz\,z^2\,\Phi(z) \over \int dz\,z\,\Phi(z)}=0.414.
\label{24}
\end{equation}
Thus using $z=\gamma u$ in Eq.\ (\ref{22}), we obtain
the scaling function $\Psi(u)$
\begin{equation}
\Psi(u)=2.1\times 10^4\,(u^2+11.65)^{-4.8}(1+25e^{-1.864u})\ .  \label{25}
\end{equation}
\begin{figure}[tbph]
\includegraphics[width=0.45\textwidth]{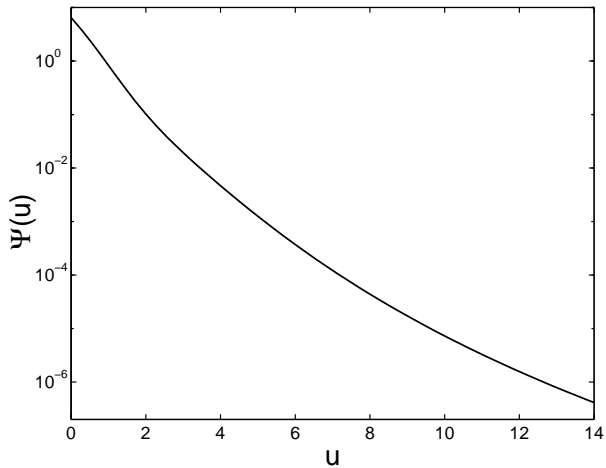}
\caption{Scaling distribution $\Psi(u)$.}
\end{figure}

In Fig.\ 5 we show $\Psi(u)$ whose shape
differs from that of $\Phi(z)$ at low $u$ because of the difference in
the power-law violating constants. $\Psi(u)$ is a universal form of
all the $P(x,N)$ shown  in Fig.\ 3.

The scaling property of $\Psi(u)$ is analogous to the KNO scaling of
the multiplicity distributions $P_m(s)$ in hadronic collisions for
$\sqrt s<200$ GeV
\cite{kno}. It was found that in terms of the scaling variable
$z=m\,/\left<m\right>$, where $m$ is the multiplicity, the KNO
function
$\psi(z)=\left<m\right>P_m(s)$ is independent of $s$. Here, we find
that
$\Psi(u)$, defined in Eq.\ (\ref{18}), is independent of centrality
when the scaling variable, $u=p_T\,/\left<p_T\right>$, is used. As
it is with KNO scaling, we have
\begin{equation}
\left<u^n\right>=\int du\,u^{n+1}\,\Psi(u)=1, \qquad\text{for}\quad n=0,1.
\label{26}
\end{equation}
The higher moments are what characterize the scaling
function, and perhaps scaling violation at some point.

Recall now that  the energy scaling distribution
found in \cite{hy} is the same as the one in Fig.\ 1 here, although the
scaling factor
$K(s)$ there is different.
Since Eq. (\ref{22}) is independent of $K(s)$ or $K(N)$,
$\Psi(u)$ is thus also the scaling distribution for any energy.

It should be emphasized that no theoretical models have been
used in any part of this investigation. The discovery of a scaling
behavior over the whole $p_T$ range that has been measured offers a
simple form of the $p_T$ distribution for dynamical models to describe
at any centrality and energy. The scaling distribution provides us
with not only a simple picture of the complex $p_T$ problem, but also
a way of quantifying the degree of degradation of the transverse
momentum in the dense medium. More importantly, the mere existence of
the scaling behavior presents a phenomenological obstacle to the realization
of the theoretical expectation that deconfinement results in an anomalous
dependence of the $p_T$ distribution on centrality. The distribution, $\Phi(z)$
or $\Psi(u)$, indicates that there is no irregularity suggestive of scaling
violation, as $N$ is varied over the whole range allowed by the Au+Au collisions.

We are grateful to W.A.\ Zajc and D.\ d'Enterria for very helpful comments. This work was
supported, in part,  by the U.\ S.\
Department of Energy under Grant No. DE-FG03-96ER40972.

\end{document}